\documentclass[conference,10pt,twocolumn,a4paper]{IEEEtran}

\IEEEoverridecommandlockouts

 \usepackage{algorithm}
 \usepackage{algorithmic}
\usepackage{amsmath,amsfonts} 
\usepackage{amssymb, amsthm}
\usepackage{graphicx}
\usepackage{subfigure}
\usepackage{booktabs}
\usepackage{multirow}
\usepackage{mathrsfs}
\usepackage{cite}
\usepackage{units}
\usepackage{cancel}
\usepackage{upgreek}
\usepackage{xcolor}
\usepackage{microtype}
\usepackage{hyperref}
\usepackage{framed}
\usepackage{algorithm}
\usepackage{balance}

\newcommand{\OPP}{{\text{OPP}}}

\newcommand{\tallm}{\overline{\bH}^\Upsilon}

\newcommand{\snr}{\varrho} 

\newtheorem{alg}{Algorithm}

\usepackage{amssymb}
\usepackage{amsfonts}
\usepackage{mathrsfs}
\usepackage{xspace}
\usepackage{bm}
\usepackage{upgreek}

\newcommand{\safemath}[2]{\newcommand{#1}{\ensuremath{#2}\xspace}}

\safemath{\bma}{\mathbf{a}}
\safemath{\bmb}{\mathbf{b}}
\safemath{\bmc}{\mathbf{c}}
\safemath{\bmd}{\mathbf{d}}
\safemath{\bme}{\mathbf{e}}
\safemath{\bmf}{\mathbf{f}}
\safemath{\bmg}{\mathbf{g}}
\safemath{\bmh}{\mathbf{h}}
\safemath{\bmi}{\mathbf{i}}
\safemath{\bmj}{\mathbf{j}}
\safemath{\bmk}{\mathbf{k}}
\safemath{\bml}{\mathbf{l}}
\safemath{\bmm}{\mathbf{m}}
\safemath{\bmn}{\mathbf{n}}
\safemath{\bmo}{\mathbf{o}}
\safemath{\bmp}{\mathbf{p}}
\safemath{\bmq}{\mathbf{q}}
\safemath{\bmr}{\mathbf{r}}
\safemath{\bms}{\mathbf{s}}
\safemath{\bmt}{\mathbf{t}}
\safemath{\bmu}{\mathbf{u}}
\safemath{\bmv}{\mathbf{v}}
\safemath{\bmw}{\mathbf{w}}
\safemath{\bmx}{\mathbf{x}}
\safemath{\bmy}{\mathbf{y}}
\safemath{\bmz}{\mathbf{z}}
\safemath{\bmzero}{\mathbf{0}}
\safemath{\bmone}{\mathbf{1}}

\bmdefine{\biad}{a}
\bmdefine{\bibd}{b}
\bmdefine{\bicd}{c}
\bmdefine{\bidd}{d}
\bmdefine{\bied}{e}
\bmdefine{\bifd}{f}
\bmdefine{\bigd}{g}
\bmdefine{\bihd}{h}
\bmdefine{\biid}{i}
\bmdefine{\bijd}{j}
\bmdefine{\bikd}{k}
\bmdefine{\bild}{l}
\bmdefine{\bimd}{m}
\bmdefine{\bind}{n}
\bmdefine{\biod}{o}
\bmdefine{\bipd}{p}
\bmdefine{\biqd}{q}
\bmdefine{\bird}{r}
\bmdefine{\bisd}{s}
\bmdefine{\bitd}{t}
\bmdefine{\biud}{u}
\bmdefine{\bivd}{v}
\bmdefine{\biwd}{w}
\bmdefine{\bixd}{x}
\bmdefine{\biyd}{y}
\bmdefine{\bizd}{z}

\bmdefine{\bixid}{\xi}
\bmdefine{\bilambdad}{\lambda}
\bmdefine{\bimud}{\mu}
\bmdefine{\bithetad}{\theta}
\bmdefine{\biphid}{\phi}
\bmdefine{\bideltad}{\delta}

\safemath{\bmia}{\biad}
\safemath{\bmib}{\bibd}
\safemath{\bmic}{\bicd}
\safemath{\bmid}{\bidd}
\safemath{\bmie}{\bied}
\safemath{\bmif}{\bifd}
\safemath{\bmig}{\bigd}
\safemath{\bmih}{\bihd}
\safemath{\bmii}{\biid}
\safemath{\bmij}{\bijd}
\safemath{\bmik}{\bikd}
\safemath{\bmil}{\bild}
\safemath{\bmim}{\bimd}
\safemath{\bmin}{\bind}
\safemath{\bmio}{\biod}
\safemath{\bmip}{\bipd}
\safemath{\bmiq}{\biqd}
\safemath{\bmir}{\bird}
\safemath{\bmis}{\bisd}
\safemath{\bmit}{\bitd}
\safemath{\bmiu}{\biud}
\safemath{\bmiv}{\bivd}
\safemath{\bmiw}{\biwd}
\safemath{\bmix}{\bixd}
\safemath{\bmiy}{\biyd}
\safemath{\bmiz}{\bizd}

\safemath{\bmxi}{\bixid}
\safemath{\bmlambda}{\bilambdad}
\safemath{\bmmu}{\bimud}
\safemath{\bmtheta}{\bithetad}
\safemath{\bmphi}{\biphid}
\safemath{\bmdelta}{\bideltad}

\safemath{\bA}{\mathbf{A}}
\safemath{\bB}{\mathbf{B}}
\safemath{\bC}{\mathbf{C}}
\safemath{\bD}{\mathbf{D}}
\safemath{\bE}{\mathbf{E}}
\safemath{\bF}{\mathbf{F}}
\safemath{\bG}{\mathbf{G}}
\safemath{\bH}{\mathbf{H}}
\safemath{\bI}{\mathbf{I}}
\safemath{\bJ}{\mathbf{J}}
\safemath{\bK}{\mathbf{K}}
\safemath{\bL}{\mathbf{L}}
\safemath{\bM}{\mathbf{M}}
\safemath{\bN}{\mathbf{N}}
\safemath{\bO}{\mathbf{O}}
\safemath{\bP}{\mathbf{P}}
\safemath{\bQ}{\mathbf{Q}}
\safemath{\bR}{\mathbf{R}}
\safemath{\bS}{\mathbf{S}}
\safemath{\bT}{\mathbf{T}}
\safemath{\bU}{\mathbf{U}}
\safemath{\bV}{\mathbf{V}}
\safemath{\bW}{\mathbf{W}}
\safemath{\bX}{\mathbf{X}}
\safemath{\bY}{\mathbf{Y}}
\safemath{\bZ}{\mathbf{Z}}

\safemath{\bZero}{\mathbf{0}}
\safemath{\bOne}{\mathbf{1}}
\safemath{\bDelta}{\mathbf{\Delta}}
\safemath{\bLambda}{\mathbf{\UpLambda}}
\safemath{\bPhi}{\mathbf{\Upphi}}
\safemath{\bSigma}{\mathbf{\Upsigma}}
\safemath{\bOmega}{\mathbf{\Upomega}}
\safemath{\bTheta}{\mathbf{\Uptheta}}

\bmdefine{\biAd}{A}
\bmdefine{\biBd}{B}
\bmdefine{\biCd}{C}
\bmdefine{\biDd}{D}
\bmdefine{\biEd}{E}
\bmdefine{\biFd}{F}
\bmdefine{\biGd}{G}
\bmdefine{\biHd}{H}
\bmdefine{\biId}{I}
\bmdefine{\biJd}{J}
\bmdefine{\biKd}{K}
\bmdefine{\biLd}{L}
\bmdefine{\biMd}{M}
\bmdefine{\biOd}{N}
\bmdefine{\biPd}{O}
\bmdefine{\biQd}{P}
\bmdefine{\biRd}{R}
\bmdefine{\biSd}{S}
\bmdefine{\biTd}{T}
\bmdefine{\biUd}{U}
\bmdefine{\biVd}{V}
\bmdefine{\biWd}{W}
\bmdefine{\biXd}{X}
\bmdefine{\biYd}{Y}
\bmdefine{\biZd}{Z}

\bmdefine{\biDelta}{\Delta}
\bmdefine{\biLambda}{\Lambda}
\bmdefine{\biPhi}{\Phi}
\bmdefine{\biSigma}{\Sigma}
\bmdefine{\biOmega}{\Omega}
\bmdefine{\biTheta}{\Theta}

\safemath{\bimA}{\biAd}
\safemath{\bimB}{\biBd}
\safemath{\bimC}{\biCd}
\safemath{\bimD}{\biDd}
\safemath{\bimE}{\biEd}
\safemath{\bimF}{\biFd}
\safemath{\bimG}{\biGd}
\safemath{\bimH}{\biHd}
\safemath{\bimI}{\biId}
\safemath{\bimJ}{\biJd}
\safemath{\bimK}{\biKd}
\safemath{\bimL}{\biLd}
\safemath{\bimM}{\biMd}
\safemath{\bimN}{\biNd}
\safemath{\bimO}{\biOd}
\safemath{\bimP}{\biPd}
\safemath{\bimQ}{\biQd}
\safemath{\bimR}{\biRd}
\safemath{\bimS}{\biSd}
\safemath{\bimT}{\biTd}
\safemath{\bimU}{\biUd}
\safemath{\bimV}{\biVd}
\safemath{\bimW}{\biWd}
\safemath{\bimX}{\biXd}
\safemath{\bimY}{\biYd}
\safemath{\bimZ}{\biZd}

\safemath{\bimDelta}{\biDelta}
\safemath{\bimLambda}{\biLambda}
\safemath{\bimPhi}{\biPhi}
\safemath{\bimSigma}{\biSigma}
\safemath{\bimOmega}{\biOmega}
\safemath{\bimTheta}{\biTheta}

\safemath{\setA}{\mathcal{A}}
\safemath{\setB}{\mathcal{B}}
\safemath{\setC}{\mathcal{C}}
\safemath{\setD}{\mathcal{D}}
\safemath{\setE}{\mathcal{E}}
\safemath{\setF}{\mathcal{F}}
\safemath{\setG}{\mathcal{G}}
\safemath{\setH}{\mathcal{H}}
\safemath{\setI}{\mathcal{I}}
\safemath{\setJ}{\mathcal{J}}
\safemath{\setK}{\mathcal{K}}
\safemath{\setL}{\mathcal{L}}
\safemath{\setM}{\mathcal{M}}
\safemath{\setN}{\mathcal{N}}
\safemath{\setO}{\mathcal{O}}
\safemath{\setP}{\mathcal{P}}
\safemath{\setQ}{\mathcal{Q}}
\safemath{\setR}{\mathcal{R}}
\safemath{\setS}{\mathcal{S}}
\safemath{\setT}{\mathcal{T}}
\safemath{\setU}{\mathcal{U}}
\safemath{\setV}{\mathcal{V}}
\safemath{\setW}{\mathcal{W}}
\safemath{\setX}{\mathcal{X}}
\safemath{\setY}{\mathcal{Y}}
\safemath{\setZ}{\mathcal{Z}}
\safemath{\emptySet}{\varnothing}

\safemath{\colA}{\mathscr{A}}
\safemath{\colB}{\mathscr{B}}
\safemath{\colC}{\mathscr{C}}
\safemath{\colD}{\mathscr{D}}
\safemath{\colE}{\mathscr{E}}
\safemath{\colF}{\mathscr{F}}
\safemath{\colG}{\mathscr{G}}
\safemath{\colH}{\mathscr{H}}
\safemath{\colI}{\mathscr{I}}
\safemath{\colJ}{\mathscr{J}}
\safemath{\colK}{\mathscr{K}}
\safemath{\colL}{\mathscr{L}}
\safemath{\colM}{\mathscr{M}}
\safemath{\colN}{\mathscr{N}}
\safemath{\colO}{\mathscr{O}}
\safemath{\colP}{\mathscr{P}}
\safemath{\colQ}{\mathscr{Q}}
\safemath{\colR}{\mathscr{R}}
\safemath{\colS}{\mathscr{S}}
\safemath{\colT}{\mathscr{T}}
\safemath{\colU}{\mathscr{U}}
\safemath{\colV}{\mathscr{V}}
\safemath{\colW}{\mathscr{W}}
\safemath{\colX}{\mathscr{X}}
\safemath{\colY}{\mathscr{Y}}
\safemath{\colZ}{\mathscr{Z}}

\safemath{\opA}{\mathbb{A}}
\safemath{\opB}{\mathbb{B}}
\safemath{\opC}{\mathbb{C}}
\safemath{\opD}{\mathbb{D}}
\safemath{\opE}{\mathbb{E}}
\safemath{\opF}{\mathbb{F}}
\safemath{\opG}{\mathbb{G}}
\safemath{\opH}{\mathbb{H}}
\safemath{\opI}{\mathbb{I}}
\safemath{\opJ}{\mathbb{J}}
\safemath{\opK}{\mathbb{K}}
\safemath{\opL}{\mathbb{L}}
\safemath{\opM}{\mathbb{M}}
\safemath{\opN}{\mathbb{N}}
\safemath{\opO}{\mathbb{O}}
\safemath{\opP}{\mathbb{P}}
\safemath{\opQ}{\mathbb{Q}}
\safemath{\opR}{\mathbb{R}}
\safemath{\opS}{\mathbb{S}}
\safemath{\opT}{\mathbb{T}}
\safemath{\opU}{\mathbb{U}}
\safemath{\opV}{\mathbb{V}}
\safemath{\opW}{\mathbb{W}}
\safemath{\opX}{\mathbb{X}}
\safemath{\opY}{\mathbb{Y}}
\safemath{\opZ}{\mathbb{Z}}
\safemath{\opZero}{\mathbb{O}}
\safemath{\identityop}{\opI}

\safemath{\veca}{\bma}
\safemath{\vecb}{\bmb}
\safemath{\vecc}{\bmc}
\safemath{\vecd}{\bmd}
\safemath{\vece}{\bme}
\safemath{\vecf}{\bmf}
\safemath{\vecg}{\bmg}
\safemath{\vech}{\bmh}
\safemath{\veci}{\bmi}
\safemath{\vecj}{\bmj}
\safemath{\veck}{\bmk}
\safemath{\vecl}{\bml}
\safemath{\vecm}{\bmm}
\safemath{\vecn}{\bmn}
\safemath{\veco}{\bmo}
\safemath{\vecp}{\bmp}
\safemath{\vecq}{\bmq}
\safemath{\vecr}{\bmr}
\safemath{\vecs}{\bms}
\safemath{\vect}{\bmt}
\safemath{\vecu}{\bmu}
\safemath{\vecv}{\bmv}
\safemath{\vecw}{\bmw}
\safemath{\vecx}{\bmx}
\safemath{\vecy}{\bmy}
\safemath{\vecz}{\bmz}

\safemath{\veczero}{\bmzero}
\safemath{\vecone}{\bmone}
\safemath{\vecxi}{\bmxi}
\safemath{\veclambda}{\bmlambda}
\safemath{\vecmu}{\bmmu}
\safemath{\vectheta}{\bmtheta}
\safemath{\vecphi}{\bmphi}
\safemath{\vecdelta}{\bmdelta}

\safemath{\matA}{\bA}
\safemath{\matB}{\bB}
\safemath{\matC}{\bC}
\safemath{\matD}{\bD}
\safemath{\matE}{\bE}
\safemath{\matF}{\bF}
\safemath{\matG}{\bG}
\safemath{\matH}{\bH}
\safemath{\matI}{\bI}
\safemath{\matJ}{\bJ}
\safemath{\matK}{\bK}
\safemath{\matL}{\bL}
\safemath{\matM}{\bM}
\safemath{\matN}{\bN}
\safemath{\matO}{\bO}
\safemath{\matP}{\bP}
\safemath{\matQ}{\bQ}
\safemath{\matR}{\bR}
\safemath{\matS}{\bS}
\safemath{\matT}{\bT}
\safemath{\matU}{\bU}
\safemath{\matV}{\bV}
\safemath{\matW}{\bW}
\safemath{\matX}{\bX}
\safemath{\matY}{\bY}
\safemath{\matZ}{\bZ}
\safemath{\matzero}{\bmzero}

\safemath{\matDelta}{\bDelta}
\safemath{\matLambda}{\bLambda}
\safemath{\matPhi}{\bPhi}
\safemath{\matSigma}{\bSigma}
\safemath{\matOmega}{\bOmega}
\safemath{\matTheta}{\bTheta}

\safemath{\matidentity}{\matI}
\safemath{\matone}{\matO}

\safemath{\rnda}{A}
\safemath{\rndb}{B}
\safemath{\rndc}{C}
\safemath{\rndd}{D}
\safemath{\rnde}{E}
\safemath{\rndf}{F}
\safemath{\rndg}{G}
\safemath{\rndh}{H}
\safemath{\rndi}{I}
\safemath{\rndj}{J}
\safemath{\rndk}{K}
\safemath{\rndl}{L}
\safemath{\rndm}{M}
\safemath{\rndn}{N}
\safemath{\rndo}{O}
\safemath{\rndp}{P}
\safemath{\rndq}{Q}
\safemath{\rndr}{R}
\safemath{\rnds}{S}
\safemath{\rndt}{T}
\safemath{\rndu}{U}
\safemath{\rndv}{V}
\safemath{\rndw}{W}
\safemath{\rndx}{X}
\safemath{\rndy}{Y}
\safemath{\rndz}{Z}

\safemath{\rveca}{\bimA}
\safemath{\rvecb}{\bimB}
\safemath{\rvecc}{\bimC}
\safemath{\rvecd}{\bimD}
\safemath{\rvece}{\bimE}
\safemath{\rvecf}{\bimF}
\safemath{\rvecg}{\bimG}
\safemath{\rvech}{\bimH}
\safemath{\rveci}{\bimI}
\safemath{\rvecj}{\bimJ}
\safemath{\rveck}{\bimK}
\safemath{\rvecl}{\bimL}
\safemath{\rvecm}{\bimM}
\safemath{\rvecn}{\bimN}
\safemath{\rveco}{\bomO}
\safemath{\rvecp}{\bimP}
\safemath{\rvecq}{\bimQ}
\safemath{\rvecr}{\bimR}
\safemath{\rvecs}{\bimS}
\safemath{\rvect}{\bimT}
\safemath{\rvecu}{\bimU}
\safemath{\rvecv}{\bimV}
\safemath{\rvecw}{\bimW}
\safemath{\rvecx}{\bimX}
\safemath{\rvecy}{\bimY}
\safemath{\rvecz}{\bimZ}

\safemath{\rvecxi}{\bmxi}
\safemath{\rveclambda}{\bmlambda}
\safemath{\rvecmu}{\bmmu}
\safemath{\rvectheta}{\bmtheta}
\safemath{\rvecphi}{\bmphi}

\safemath{\rmatA}{\bimA}
\safemath{\rmatB}{\bimB}
\safemath{\rmatC}{\bimC}
\safemath{\rmatD}{\bimD}
\safemath{\rmatE}{\bimE}
\safemath{\rmatF}{\bimF}
\safemath{\rmatG}{\bimG}
\safemath{\rmatH}{\bimH}
\safemath{\rmatI}{\bimI}
\safemath{\rmatJ}{\bimJ}
\safemath{\rmatK}{\bimK}
\safemath{\rmatL}{\bimL}
\safemath{\rmatM}{\bimM}
\safemath{\rmatN}{\bimN}
\safemath{\rmatO}{\bimO}
\safemath{\rmatP}{\bimP}
\safemath{\rmatQ}{\bimQ}
\safemath{\rmatR}{\bimR}
\safemath{\rmatS}{\bimS}
\safemath{\rmatT}{\bimT}
\safemath{\rmatU}{\bimU}
\safemath{\rmatV}{\bimV}
\safemath{\rmatW}{\bimW}
\safemath{\rmatX}{\bimX}
\safemath{\rmatY}{\bimY}
\safemath{\rmatZ}{\bimZ}

\safemath{\rmatDelta}{\bimDelta}
\safemath{\rmatLambda}{\bimLambda}
\safemath{\rmatPhi}{\bimPhi}
\safemath{\rmatSigma}{\bimSigma}
\safemath{\rmatOmega}{\bimOmega}
\safemath{\rmatTheta}{\bimTheta}

\usepackage{amssymb}
\usepackage{amsfonts}
\usepackage{mathrsfs}
\usepackage{xspace}
\usepackage{bm}
\usepackage{fancyref}
\usepackage{textcomp}

\usepackage{multirow}
\usepackage{stmaryrd}

\newenvironment{textbmatrix}{	\setlength{\arraycolsep}{2.5pt}%
								\big[\begin{matrix}}{\end{matrix}\big]%
								\raisebox{0.08ex}{\vphantom{M}}}

\def\be{\begin{equation}}
\def\ee{\end{equation}}
\def\een{\nonumber \end{equation}}
\def\mat{\begin{bmatrix}}
\def\emat{\end{bmatrix}}
\def\btm{\begin{textbmatrix}}
\def\etm{\end{textbmatrix}}

\def\ba#1\ea{\begin{align}#1\end{align}}
\def\bas#1\eas{\begin{align*}#1\end{align*}}
\def\bs#1\es{\begin{split}#1\end{split}} 
\def\bg#1\eg{\begin{gather}#1\end{gather}}
\def\bml#1\eml{\begin{multline}#1\end{multline}}
\def\bi#1\ei{\begin{itemize}#1\end{itemize}}

\newcommand{\lefto}{\mathopen{}\left}

\DeclareMathOperator*{\argmin}{arg\;min}		

\newcommand{\vecnorm}[1]{\lefto\lVert#1\right\rVert}		

\safemath{\dirac}{\delta}					
\safemath{\krond}{\dirac}					

\safemath{\upto}{\uparrow}
\safemath{\downto}{\downarrow}
\safemath{\iu}{j}							
\safemath{\ev}{\lambda}						
\safemath{\hilseqspace}{l^{2}}				
\newcommand{\banachfunspace}[1]{\setL^{#1}}	
\safemath{\hilfunspace}{\banachfunspace{2}}	

\safemath{\SNR}{\textsf{SNR}} 				
\safemath{\PAR}{\textsf{PAR}} 				
\safemath{\No}{N_0}							
\safemath{\Es}{E_s}							
\safemath{\Eb}{E_b}							
\safemath{\EbNo}{\frac{\Eb}{\No}}
\safemath{\EsNo}{\frac{\Es}{\No}}

\DeclareMathOperator{\CHop}{\ensuremath{\opH}} 
\safemath{\tvir}{\rndh_{\CHop}}				
\safemath{\tvtf}{\rndl_{\CHop}}				
\safemath{\spf}{\rnds_{\CHop}}				
\safemath{\bff}{H_{\CHop}}					

\safemath{\ircf}{r_{h}}						
\safemath{\tftvcf}{r_{s}}					
\safemath{\tfcf}{r_{l}}						
\safemath{\bfcf}{r_{H}}						

\safemath{\tcorr}{c_h}						
\safemath{\scf}{c_{s}}						
\safemath{\tfcorr}{c_{l}}					
\safemath{\fcorr}{c_{H}}						

\safemath{\mi}{I}							
\safemath{\capacity}{C}						

\safemath{\normal}{\mathcal{N}}			
\safemath{\jpg}{\mathcal{CN}}			
\safemath{\mchain}{\leftrightarrow}		

\safemath{\dB}{\,\mathrm{dB}}
\safemath{\dBm}{\,\mathrm{dBm}}
\safemath{\Hz}{\,\mathrm{Hz}}
\safemath{\kHz}{\,\mathrm{kHz}}
\safemath{\MHz}{\,\mathrm{MHz}}
\safemath{\GHz}{\,\mathrm{GHz}}
\safemath{\s}{\,\mathrm{s}}
\safemath{\ms}{\,\mathrm{ms}}
\safemath{\mus}{\,\mathrm{\text{\textmu}s}}
\safemath{\ns}{\,\mathrm{ns}}
\safemath{\ps}{\,\mathrm{ps}}
\safemath{\meter}{\,\mathrm{m}}
\safemath{\mm}{\,\mathrm{mm}}
\safemath{\cm}{\,\mathrm{cm}}
\safemath{\m}{\,\mathrm{m}}
\safemath{\W}{\,\mathrm{W}}
\safemath{\mW}{\, \mathrm{mW}}
\safemath{\J}{\,\mathrm{J}}
\safemath{\K}{\,\mathrm{K}}
\safemath{\bit}{\,\mathrm{bit}}
\safemath{\nat}{\,\mathrm{nat}}

\safemath{\define}{\triangleq}			

\safemath{\equivalent}{\sim}
\safemath{\distas}{\sim}					
\safemath{\sdiff}{\Delta}				

\safemath{\reals}{\mathbb{R}}
\safemath{\positivereals}{\reals_{+}}
\safemath{\integers}{\mathbb{Z}}
\safemath{\posint}{\integers_{+}}
\safemath{\naturals}{\mathbb{N}}
\safemath{\posnaturals}{\naturals_{+}}
\safemath{\complexset}{\mathbb{C}}
\safemath{\rationals}{\mathbb{Q}}

\newcommand*{\fancyrefapplabelprefix}{app}		
\newcommand*{\fancyrefthmlabelprefix}{thm}		
\newcommand*{\fancyreflemlabelprefix}{lem}		
\newcommand*{\fancyrefcorlabelprefix}{cor}		
\newcommand*{\fancyrefdeflabelprefix}{def}		
\newcommand*{\fancyrefproplabelprefix}{prop}	
\newcommand*{\fancyrefobslabelprefix}{obs}		
\newcommand*{\fancyrefalglabelprefix}{alg}		
\newcommand*{\fancyrefasmlabelprefix}{asm}	    
\newcommand*{\fancyreftbllabelprefix}{tbl}	    

\frefformat{vario}{\fancyrefseclabelprefix}{Sec.~#1}
\frefformat{vario}{\fancyrefthmlabelprefix}{Thm.~#1}
\frefformat{vario}{\fancyreflemlabelprefix}{Lem.~#1}
\frefformat{vario}{\fancyrefcorlabelprefix}{Cor.~#1}
\frefformat{vario}{\fancyrefdeflabelprefix}{Def.~#1}
\frefformat{vario}{\fancyrefobslabelprefix}{Obs.~#1}
\frefformat{vario}{\fancyrefasmlabelprefix}{Ass.~#1}
\frefformat{vario}{\fancyreffiglabelprefix}{Fig.~#1}
\frefformat{vario}{\fancyrefapplabelprefix}{App.~#1} 
\frefformat{vario}{\fancyrefproplabelprefix}{Prop.~#1}
\frefformat{vario}{\fancyrefalglabelprefix}{Alg.~#1}
\frefformat{vario}{\fancyrefeqlabelprefix}{(#1)}
\frefformat{vario}{\fancyreftbllabelprefix}{Tbl.~#1}

\safemath{\dictab}{[\,\dicta\,\,\dictb\,]}

\safemath{\ysig}{\bmy}
\safemath{\ysighat}{\hat{\ysig}}
\safemath{\ysigdim}{M}
\safemath{\xsig}{\bmx}
\safemath{\xsigdim}{N}
\safemath{\nx}{n_x}
\safemath{\zsig}{\bmz}
\safemath{\zsigdim}{\ysigdim}
\safemath{\rsig}{\bmr}
\safemath{\Adict}{\bA}
\safemath{\Adicttilde}{\widetilde{\Adict}}
\safemath{\Adictdim}{\outputdim\times\xsigdim}
\safemath{\avec}{\bma}
\safemath{\avectilde}{\tilde{\avec}}
\safemath{\Bdict}{\bB}
\safemath{\Bdicttilde}{\widetilde{\Bdict}}
\safemath{\Cdict}{\bC}
\safemath{\cvec}{\bmc}
\safemath{\Ddict}{\bD}
\safemath{\Ddictdim}{\ysigdim\times\xsigdim}
\safemath{\dvec}{\bmd}
\safemath{\Ddicttilde}{\widetilde{\bD}}
\safemath{\Bonb}{\bB}
\safemath{\bvec}{\bmb}
\safemath{\Bonbdim}{\ysigdim\times\ysigdim}
\safemath{\noise}{\bmn}
\safemath{\noisedim}{\ysigim}
\safemath{\err}{\bme}
\safemath{\errdim}{\ysigdim}
\safemath{\errset}{\setE}
\safemath{\nerr}{n_e}
\safemath{\delop}{\bP_\errset}
\safemath{\delopc}{\bP_{{\errset}^c}}

\safemath{\cplxi}{\imath}
\safemath{\cplxj}{\jmath}

\safemath{\dict}{\matD}
\safemath{\inputdim}{N}		
\safemath{\outputdim}{M}		
\safemath{\sparsity}{S}	
\safemath{\inputdimA}{{N_a}}	
\safemath{\inputdimB}{{N_b}}	
\safemath{\elemA}{{n_a}}	
\safemath{\elemB}{{n_b}}	
\safemath{\resA}{\matR_a}	
\safemath{\resB}{\matR_b}	
\safemath{\subD}{\matS} 
\safemath{\subA}{\matS_a} 
\safemath{\subB}{\matS_b} 
\safemath{\dicta}{\matA} 	
\safemath{\dictb}{\matB} 	
\safemath{\hollowS}{H}
\safemath{\hollowA}{H_a}
\safemath{\hollowB}{H_b}
\safemath{\cross}{Z}
\safemath{\coh}{\mu_d}			
\safemath{\coha}{\mu_a}			
\safemath{\cohb}{\mu_b}			
\safemath{\mubs}{\nu}	
\safemath{\cohm}{\mu_m} 
\safemath{\dictset}{\setD}	
\safemath{\dictsetp}{\dictset(\coh,\coha,\cohb)}	
\safemath{\dictsetgen}{\dictset_\text{gen}}
\safemath{\dictsetgenp}{\dictsetgen(\coh)}
\safemath{\dictsetonb}{\dictset_\text{onb}}
\safemath{\dictsetonbp}{\dictsetonb(\coh)}

\safemath{\leftside}{U}
\safemath{\rightsideA}{R_a}
\safemath{\rightsideB}{R_b}

\safemath{\indexS}{\setI_S} 

\safemath{\na}{n_a}			
\safemath{\nb}{n_b}			
\safemath{\coeffa}{p_i}	
\safemath{\coeffb}{q_j}	
\safemath{\seta}{\setP}		
\safemath{\setb}{\setQ}     
\safemath{\setw}{\setW}	
\safemath{\setz}{\setZ}	
\safemath{\cola}{\veca}		
\safemath{\colb}{\vecb}		
\safemath{\cold}{\vecd}		
\safemath{\inputvec}{\vecx} 	
\safemath{\error}{\vece}	
\safemath{\noiseout}{\vecz} 	
\safemath{\inputvecel}{x}
\safemath{\inputveca}{\vecx_a}
\safemath{\inputvecb}{\vecx_b}
\safemath{\outputvec}{\vecy}	
\safemath{\lambdamin}{\lambda_{\mathrm{min}}}

\safemath{\elltwo}{\ell_2}
\safemath{\ellone}{\ell_1}
\safemath{\ellzero}{\ell_0}
\safemath{\ellinf}{\ell_\infty}
\safemath{\ellinftilde}{\ell_{\widetilde\infty}}
\safemath{\licard}{Z(\coh,\coha,\cohb)}
\safemath{\xsol}{\hat{x}}
\safemath{\xbord}{x_b}		
\safemath{\xstat}{x_s}		
\safemath{\xstatLone}{\tilde{x}_s}
\safemath{\order}{\mathcal{O}} 
\safemath{\scales}{\Theta} 
\safemath{\ones}{\mathbf{1}} 
\safemath{\zeroes}{\mathbf{0}} 
\safemath{\thlone}{\kappa(\coh,\cohb)} 
\safemath{\constoneA}{\delta} 
\safemath{\constoneB}{\epsilon} 
\safemath{\nlarge}{L}				   
\safemath{\sumlarge}{S_\nlarge}
\safemath{\maxlarger}{P_\nlarge}	   
\safemath{\Pzero}{\textrm{P0}}	
\safemath{\Pone}{\textrm{P1}}
\safemath{\vecfir}{\vecw}			 
\safemath{\vecsec}{\vecz}
\safemath{\elvecfir}{w}              
\safemath{\elvecsec}{z}				 
\safemath{\nlargefir}{n}
\safemath{\normout}{\gamma}
\safemath{\auxfun}{h}
\safemath{\supp}{\textrm{supp}}

\safemath{\indexa}{\ell}
\safemath{\indexb}{r}
\safemath{\indexc}{i}
\safemath{\indexd}{j}

\safemath{\project}{P}

\renewcommand{\bml}{\ensuremath{\boldsymbol \ell}}

\definecolor{brightpink}{rgb}{1.0, 0.0, 0.5}

\begin{document}
\title{VLSI Design of a  3-bit Constant-Modulus Precoder \\ for Massive MU-MIMO} 
\author{
\IEEEauthorblockN{Oscar Casta\~neda$^{1}$, Sven Jacobsson$^\text{1,2,3}$, Giuseppe Durisi$^\text{2}$, Tom Goldstein$^4$, and Christoph Studer$^1$} \\ \vspace{-0.1cm}          \IEEEauthorblockA{$^\text{1}$Cornell University, Ithaca, NY; \url{oc66@cornell.edu}; \url{studer@cornell.edu}; web: \url{http://vip.ece.cornell.edu}} 
\IEEEauthorblockA{$^\text{2}$Ericsson Research, Gothenburg, Sweden; \url{sven.jacobsson@ericsson.com}} 
\IEEEauthorblockA{$^\text{3}$Chalmers University of Technology, Gothenburg, Sweden; \url{durisi@chalmers.se}} 
\IEEEauthorblockA{$^\text{4}$University of Maryland, College Park, MD; \url{tomg@cs.umd.edu}}
\vspace{-0.1 in}
\thanks{The precoding algorithm and architecture proposed in this paper builds upon the one proposed in~\cite{castaneda2017cxpo} for C2PO; in contrast to these results, the present paper uses a modified architecture with a more sophisticated projection unit.}
\thanks{A MATLAB simulator for the precoder proposed in this paper is available on GitHub: \url{https://github.com/quantizedmassivemimo/3bit_CM_precoding}.}
\thanks{OC and CS were supported in part by Xilinx Inc.~and by the US NSF under grants ECCS-1408006, CCF-1535897, CAREER CCF-1652065, and CNS-1717559. The work of SJ and GD was supported in part by the Swedish Foundation for Strategic Research under grant ID14-0022, and by the Swedish Governmental Agency for Innovation Systems (VINNOVA) within the competence center ChaseOn. SJ's research visit at Cornell was sponsored in part by Cornell's College of Engineering. TG was supported by the US NSF under grant CCF-1535902 and by the US ONR under grant N00014-15-1-2676.}
}
\maketitle

\begin{abstract}
Fifth-generation (5G) cellular systems will build on massive multi-user (MU) multiple-input multiple-output (MIMO) technology to attain high spectral efficiency.
However, having hundreds of antennas and radio-frequency (RF) chains at the base station~(BS) entails prohibitively high hardware costs and power consumption. 
%
%
This paper proposes a novel nonlinear precoding algorithm for the massive MU-MIMO downlink in which each RF chain contains an 8-phase (3-bit) constant-modulus transmitter, enabling the use of low-cost and power-efficient analog hardware.
We present a high-throughput VLSI architecture and show implementation results on a Xilinx \mbox{Virtex-7} FPGA.  
Compared to a recently-reported nonlinear precoder for BS designs that use two $\boldsymbol1$-bit digital-to-analog converters per RF chain, our design enables up to $\boldsymbol{3.75}$\,dB transmit power reduction at no more than a 2.7$\boldsymbol\times$ increase in FPGA resources.
\end{abstract}



\vspace{-0.13cm}

\section{Introduction}
\label{sec:intro}

Fifth-generation (5G) cellular communication systems are widely expected to rely on massive  multi-user (MU) multiple-input multiple-output (MIMO) technology to achieve significant improvements in spectral efficiency compared to existing small-scale MIMO systems~\cite{rusek14a, larsson14a, lu14a}.
MU-MIMO equips the base station~(BS) with hundreds of antennas and radio-frequency~(RF) chains, enabling one to simultaneously serve tens of user equipments (UEs) in the same time-frequency resource via fine-grained beamforming. 
Unfortunately, scaling conventional multi-antenna BS architectures (that use high-precision RF chains) to BSs with hundreds of antenna elements entails a significant increase in system costs and circuit power consumption. Hence, to make massive MU-MIMO systems inexpensive and power-efficient, novel BS architectures and suitable baseband-processing algorithms are~necessary. 

\subsubsection*{Low-Precision BS Architechures}
The use of low-precision digital-to-analog-converters (DACs) at the BS in the massive MU-MIMO downlink enables significant reductions in terms of system costs and circuit power consumption.
%
%
The key challenge with such low-precision BS architectures is to maintain high spectral efficiency, which requires sophisticated baseband-processing algorithms.
While linear precoders, e.g., maximal-ratio transmission (MRT) and zero-forcing (ZF), followed by quantization exhibit low complexity~\cite{mezghani09c,saxena16b,li17a}, sophisticated nonlinear precoders can achieve superior performance, especially for the extreme case of using only a pair of 1-bit DACs per RF chain~\cite{jacobsson17d, jacobsson16b,jedda16a,tirkkonen17a,swindlehurst17a,castaneda17b}. 
Recently, reference~\cite{castaneda2017cxpo} presented VLSI designs of nonlinear precoders for systems that use a pair of \mbox{$1$-bit} DACs per RF chain, which demonstrates that  nonlinear precoding is feasible in practice for such low-precision BS~architectures.

The use of 1-bit DACs at the BS ensures that the precoded signal has constant-modulus (CM), i.e., the precoded signal's amplitude is equal on all antennas and constant over time, which enables the use of low-cost and power-efficient analog circuitry, such as nonlinear power amplifiers.
Recently, nonlinear precoders for 8-phase (3-bit) CM transmitters, i.e., the setup considered in this work, were proposed in~\cite{noll17a, jacobsson17f}.
It remains, however, an open question whether the algorithms proposed in~\cite{noll17a, jacobsson17f} can be implemented efficiently in hardware.

\subsubsection*{Contributions}
This paper develops a novel nonlinear precoding algorithm in which each RF chain contains an 8-phase (3-bit) CM transmitter that enables efficient analog circuitry while surpassing the error-rate performance of systems that use a pair of $1$-bit DACs (i.e., 4 phases) per RF chain.
We propose a nonconvex algorithm to solve the associated 8-phase ($3$-bit) CM precoding problem in an efficient manner, and  we develop a VLSI architecture that uses a fast matrix-vector multiplication engine based on Cannon's algorithm~\cite{CannonThesis}.
We show Xilinx Virtex-7 FPGA implementation results and provide a comparison with the C2PO precoder proposed in~\cite{castaneda2017cxpo}. 

 

\section{System Model and CM Precoding}
\label{sec:system}

\setlength{\textfloatsep}{10pt}
\begin{figure}[t]
\centering
\includegraphics[width=0.95\columnwidth]{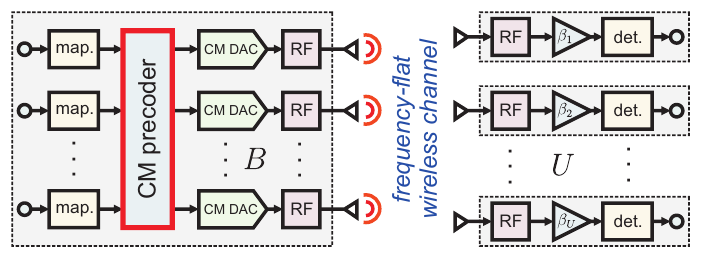}
\vspace{-0.3cm}
 \caption{Overview of the considered massive MU-MIMO downlink system with CM DACs. Left: $B$ antenna massive MU-MIMO BS containing a CM precoder that mitigates multi-user interference and quantization artifacts in the CM DACs; Right: $U$ single-antenna UEs. } 
\label{fig:system_overview}
\end{figure}

\subsection{System Model}

We consider the single-cell, narrowband massive MU-MIMO downlink system shown in \fref{fig:system_overview}. Here, the BS, which is equipped with $B$ antennas, serves $U\leq B$ single-antenna UEs. 
The narrowband downlink channel is modeled by $\vecy = \matH \vecx + \vecn$, where $\vecy = [y_1,\dots,y_U]^T\in\complexset^U$ contains the received signals at all UEs,  $\matH \in \opC^{U \times B}$ is the channel matrix (which we assume is known to the BS),~$\bmn\in\complexset^U$ models i.i.d.\ circularly-symmetric complex Gaussian noise with variance~$N_0$ per complex~entry, and $\bmx\in\setX^B$ is the so-called \emph{precoded vector}, where~$\setX$ is the transmit alphabet. 
In this work, we require that $\setX$ has finite cardinality and that the entries of $\setX$ have CM.
%
%
Specifically, the CM alphabet is $\setX=\{\exp(j2\pi p/P)\mid p=0,\ldots,P-1\}$ where~$P$ denotes the number of phases and $\log_2(P)$ the number of bits per RF chain.
The CM constraint ensures that $\vecnorm{\vecx}^2_2 = B$.
%

%

\subsection{Constant-Modulus (CM) Precoding}

The precoder at the BS maps the symbol vector $\vecs = [s_1, \dots, s_U]^T$ into the precoded vector $\vecx \in \setX^B$. Here, $s_u \in \setO$ is the constellation point intended for the $u$th UE~($u=1,\ldots,U$), where~$\setO$ is the constellation set (e.g., 16-QAM).
We assume that each UE $u=1,\ldots,U$ rescales its received signal~$y_u$ by a factor $\beta_u \in \complexset$ to compute an estimate~$\hat{s}_u = \beta_u y_u$ of the transmitted symbol $s_u$.
%
%
%
Nonlinear precoders that minimize the mean-squared error~(MSE) between the transmitted and the estimated symbols solve the following optimal precoding problem~(OPP)~\cite{castaneda2017cxpo}: 
\begin{align*}
\text{(\OPP)} \quad 
\{\hat\vecx,\hat\beta\} = \underset{\bmx \in \setX^{B}\!,\, \beta \in \complexset}{\text{arg\,min}} \,\, \vecnorm{\bms - \beta \matH\bmx}^2_2 + |\beta|^2 U \No.
\end{align*}
Here, we assume that $\beta=\beta_u$ for $u = 1, \ldots, U$; as shown in~\cite{jacobsson16b}, the UEs are able to accurately learn~$\hat{\beta}$.
%
%
For systems that use a pair of $1$-bit DACs per RF chain ($P=4$~phases), methods that solve (OPP) approximately using convex~\cite{jacobsson17d,jacobsson16b}  and nonconvex~\cite{castaneda2017cxpo} relaxation have been  proposed recently. In what follows, we present a novel precoder specifically designed for CM transmitters with $3$ bits per RF chain ($P=8$ phases), which enables significant error-rate performance improvements compared to systems with $2$ bits per RF chain ($P=4$ phases), without requiring complex RF circuitry.


%
%

\section{C3PO: Constant-modulus 3-bit PrecOding}
\label{sec:algo}

\subsection{Relaxing the Problem (\OPP{})}
\begin{figure}[tp]
\centering
\includegraphics[width=0.975\columnwidth]{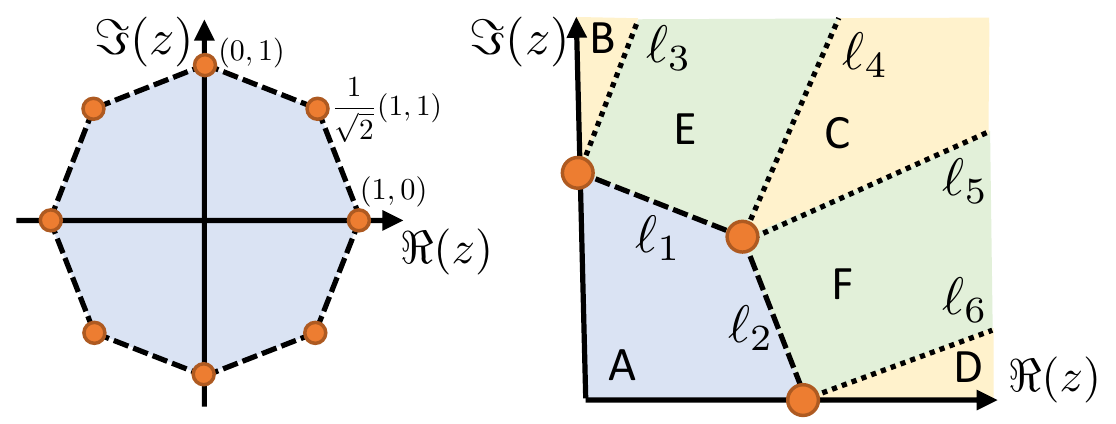}
\vspace{-0.19cm}
\caption{Left: 8-phase CM alphabet (convex polytope in blue); right: projection regions within the first quadrant of the 8-phase CM alphabet.}
\label{fig:8p_regions}
\end{figure}

To find an approximate solution to (\OPP{}) via methods that can be implemented efficiently, we perform the following approximations. First, we let $\No\to0$, i.e., we assume that the system operates in the high-SNR regime. Then, we use the following approximation~\cite[Eq.~(2)]{castaneda2017cxpo}:
\begin{align*} 
\underset{\bmx\in\setX^B}{\text{min}} \underset{\beta\in\complexset}{\text{min}}\, \vecnorm{\vecs - \beta \matH\vecx}^2_2 \approx 
\underset{\bmx\in\setX^B}{\text{min}}\underset{\alpha\in\complexset}{\text{min}}\,\vecnorm{\alpha\vecs -  \matH\vecx}^2_2.
\end{align*}
%
These two approximations result in the following problem:
\begin{align*}
\text{(\OPP$^*$)} \qquad 
\{\hat{\bmx}, \hat\alpha\} = \argmin_{\bmx \in \setX^{B},\, \alpha \in \complexset}  \vecnorm{\alpha{\vecs} -  {\matH}{\vecx}}^2_2.
\end{align*}
We next compute $\hat{\alpha}$ by minimizing the objective function of $\text{(\OPP$^*$)}$, which results in
$\hat{\alpha}=  {{\bms}^H{\bH}{\bmx}}/{\|{\bms}\|^2_2}$.
Substituting $\hat{\alpha}$ in $\text{(\OPP$^*$)}$ yields $\vecnorm{\hat\alpha(\bmx){\vecs} -  {\matH}{\vecx}}^2_2 = \vecnorm{\bA\vecx}^2_2$  with $\bA =  \bQ \bH$ and $\bQ = \bI_U-{\bms\bms^H}/{\|\bms\|^2_2}$.
Hence, we can simplify $\text{(\OPP$^*$)}$ as
\begin{align*}
\text{(\OPP$^{**}$)} \qquad 
\hat{\bmx} = \argmin_{\bmx \in \setX^{B}} \textstyle \frac{1}{2}\vecnorm{\bA\vecx}^2_2.
\end{align*}
The factor ${1}/{2}$ does not affect the solution of \text{(\OPP$^{**}$)}.
We now replace the finite-phase constraint $\bmx \in \setX^{B}$ by the convex polytope surrounding the points~$\setX=\{\mathsf{x}_p\}^{P}_{p=1}$ given by
\begin{align*}
\textstyle  \setB = \Big\{\sum_{p=1}^{P}\alpha_p \mathsf{x}_p \mid (\alpha_p\geq0,\forall p) \wedge \sum_{p=1}^{P}\alpha_p=1 \Big\}.
\end{align*}
For 3-bit CM precoding, the boundary of the convex polytope~$\setB$ is a regular octagon (see~\fref{fig:8p_regions}).  
Unfortunately, solving $\text{(\OPP$^{**}$)}$ over the relaxed set $\bmx \in \setB^{B}$ yields the all-zeros vector. 
We therefore attempt to solve the following modified problem via forward-backward splitting (FBS)~\cite{GSB14,BT09,goldstein2010high}:
\begin{align}
\hat{\bmx} = \argmin_{\bmx \in \setB^{B}} \,\, \textstyle \frac{1}{2} \vecnorm{\bA\vecx}^2_2    -  \frac{\delta}{2} \|\bmx\|_2^2, \label{eq:directform}
\end{align}
where the concave regularizer $- \frac{\delta}{2} \|\bmx\|_2^2$ with $\delta>0$ forces the solution $\hat\bmx$ to lie at the boundary of the convex polytope $\setB^{B}$. As the problem in \fref{eq:directform} is nonconvex, FBS is not guaranteed to converge to an optimal solution. Nevertheless, the algorithm proposed exhibits good empirical performance (see \fref{sec:architecture}).

\subsection{The C3PO Algorithm}
FBS is an efficient numerical method to solve convex optimization problems whose objective function can be decomposed as $f(\bmx)+g(\bmx)$, where the function $f$ is smooth and convex, and the function $g$ is convex but not necessarily smooth or bounded. FBS consists of the following iteration~\cite{BT09,GSB14}:
\begin{align*}
\bmx^{(t+1)}  \!=\! \text{prox}_g\!\big( \bmz^{(t+1)}  ; \tau^{(t)}  \big)\! \text{ with }
\bmz^{(t+1)}  \!=\! \bmx^{(t)} \!-\! \tau^{(t)} \nabla f(\bmx^{(t)}) 
\end{align*}
for $t=1,2,\ldots,t_\text{max}$ or until convergence. Here, the sequence $\{\tau^{(t)}>0\}$ contains suitably chosen step-size parameters and $\nabla f(\bmx)$ is the gradient of the smooth function $f$, and the so-called \emph{proximal operator} for the function $g$ is defined by~\cite{parikh2014proximal}
\begin{align*}
\text{prox}_g \!\left( \bmz  ; \tau  \right) = \argmin_{\bmx \in \opC^B} \textstyle \left\{ \tau g(\bmx)+\frac{1}{2}\|\bmx-\bmz\|_2^2 \right\}\!.
\end{align*}
To approximately solve \fref{eq:directform} using FBS, we set 
\begin{align*}
f(\bmx)= \textstyle \frac{1}{2}\vecnorm{\bA\vecx}^2_2 \,\text{ and }\,  g(\bmx) =  \chi\!\left(\bmx \in \setB^{B}\right)- \frac{\delta}{2}\|\bmx\|_2^2 , 
\end{align*}
where $\chi$ is a characteristic function that is zero if $\bmx \in \setB^{B}$ and infinity otherwise. For these choices, the gradient is given by  $\nabla f(\bmx)=\bA^H\bA\vecx$ and the proximal operator is detailed in \fref{sec:proximal}. Furthermore, we use a constant step size $\tau=\tau^{(t)}$.
The resulting algorithm is as follows: \\[-0.6cm]
\begin{oframed}
\vspace{-0.3cm}
\begin{alg}[C3PO] \label{alg:C3PO} 
Initialize \mbox{$\bmx^{(1)}=\bH^H\bms$} and fix the parameters $\delta$ and $\tau$ so that $\tau\delta < 1$.
Then, for every iteration $t=1,2,\ldots,t_\text{max}$ compute: 
\begin{align}
\bmz^{(t+1)} &= \bmx^{(t)} - \tau \bA^H\bA \bmx^{(t)}  \label{eq:faststep1} \\
\bmx^{(t+1)} & = \mathrm{prox}_g ( {\bmz^{(t+1)}};\tau ). \label{eq:faststep2} 
\end{align}
The $\mathrm{prox}_g$ operator is applied element-wise to $\bmz^{(t+1)}$ and detailed in \fref{sec:prox8op}. In the last iteration $t_\text{max}$, the output $\bmx^{(t_\text{max}+1)}$ is quantized to the 3-bit CM alphabet $\setX^B$.
\end{alg}
\vspace{-0.2cm}
\end{oframed}

\vspace{-0.2cm}
The most costly operation of C3PO is {the} matrix-vector product in step~\fref{eq:faststep1}, which we compute as:
$\bA^H\bA 
= \bH^H\bH - \vecv\vecv^H 
= {\tallm} \overline{\bH}$,
where $\bmv=\bH^H\bms/\|\bms\|_2$ is a normalized version of the {MRT} vector; {the augmented matrices $\overline{\bH}=[\bH ; \bmv^H ]$ and $\tallm=[\bH^H , -\bmv ]$ are of dimension $(U+1)\times B$ and $B\times (U+1)$, respectively.} Then, step~\fref{eq:faststep1} is rewritten as follows:
\begin{align}
\label{eq:faststep1-simple}
\bmz^{(t+1)} = \bmx^{(t)} - \tau {\tallm}\overline{\bH} \bmx^{(t)}.
\end{align}

\subsection{Proximal Operator for 3-Bit CM Precoding}
\label{sec:prox8op} 
\label{sec:proximal}
The proximal operator in~\eqref{eq:faststep2} reduces to $\mathrm{prox}_g ( {\bmz};\tau ) = \mathrm{proj} (\frac{1}{1-\tau\gamma} \bmz )$, where $\mathrm{proj} (\cdot)$ projects each element of the argument to the closest point in the polytope $\setB$.
For 3-bit CM precoding, the polytope is a regular octagon. Projecting a scalar $z \in \opC$ onto an octagon is nontrivial so we focus on the first quadrant of the complex plane (see \fref{fig:8p_regions}).
If $z$ is inside the octagon (in region \textsf{A}), then it remains there; if $z$ is in the regions \textsf{B}, \textsf{C}, or \textsf{D}, then it will be mapped to $j$, $\frac{1}{\sqrt{2}}(1+j)$, or $1$, respectively; if $z$ is in the regions \textsf{E} or~\textsf{F}, then it will be mapped to the closest point on the lines $\ell_1$ or $\ell_2$, respectively.
To determine in which of the six regions \textsf{A}--\textsf{F} the argument is located, we use the equations for the lines that separate them:
\begin{align*}
\ell_1:~&\Im(z)=\textstyle(1-\sqrt{2})\Re(z)+1,\\
\ell_3:~&\Im(z)=\textstyle\frac{1}{\sqrt{2}-1}\Re(z)+1,
~~~~\ell_4:~\Im(z) =  \textstyle \frac{1}{\sqrt{2}-1}\Re(z)-1.
\end{align*}
The equations for the lines $\ell_2$, $\ell_5$, and $\ell_6$ are identical to the ones of $\ell_1$, $\ell_4$, and $\ell_3$, but with $\Im(z)$ and $\Re(z)$ exchanged.
Using these equations, we can project~$z$ onto the set $\setB$. 

%
\begin{figure}[tp]
\centering
\vspace{-0.2cm}
\includegraphics[width=0.9\columnwidth]{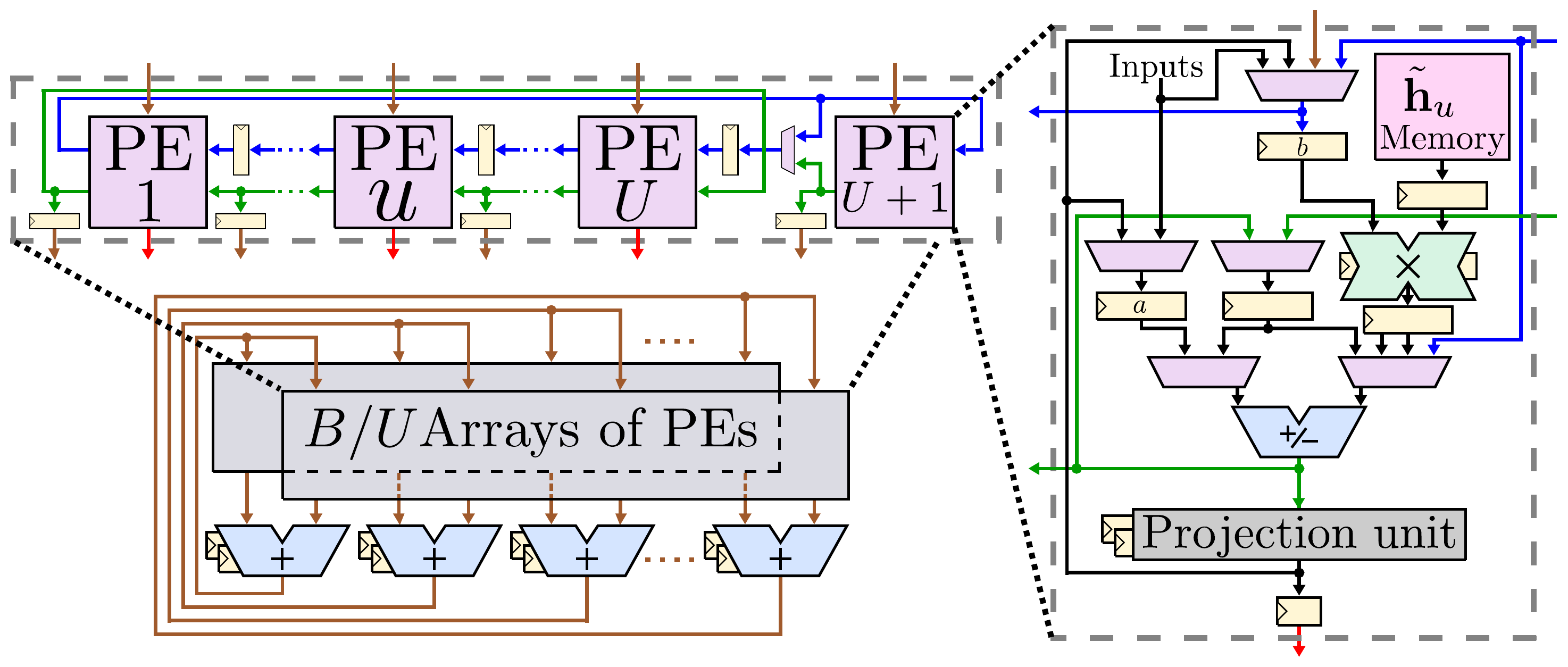}
\vspace{-0.2cm}
\caption{High-level block diagram of the  VLSI architecture for C3PO. We use $B/U$ linear arrays, each consisting of $U+1$ processing elements~(PEs).
}
\label{fig:C3POarch}
\end{figure}
%


\begin{figure*}[tp]
\centering
\subfigure[$B=32$, $U=16$, and BPSK.]{\includegraphics[width=.31\textwidth]{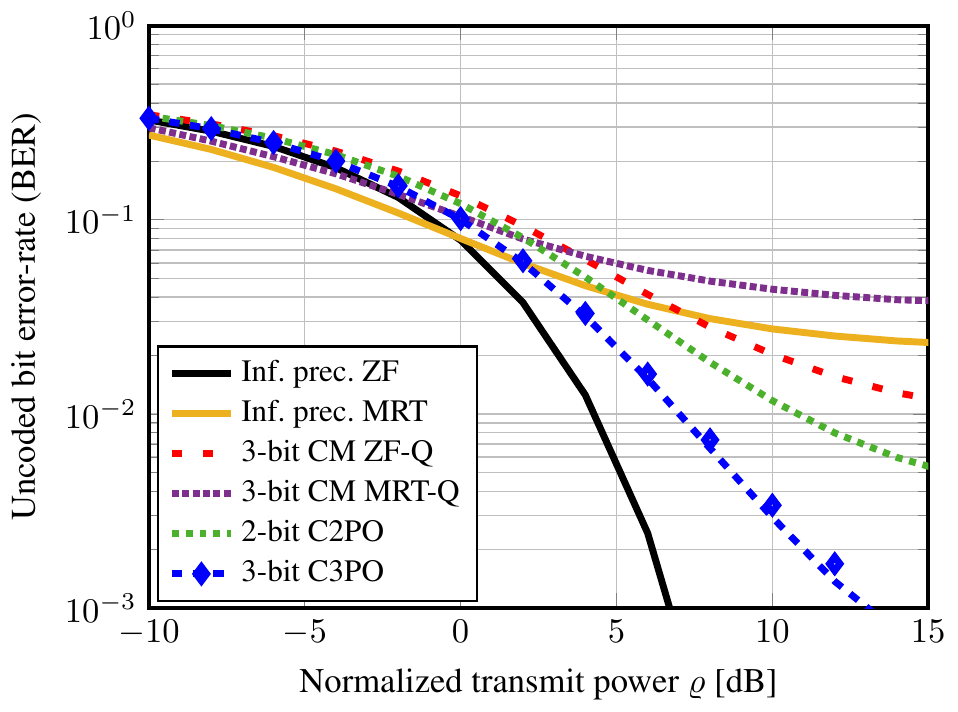}\label{fig:BER_32x16_BPSK}}
\quad
\subfigure[$B=256$, $U=16$, and 16-QAM.]{\includegraphics[width=.31\textwidth]{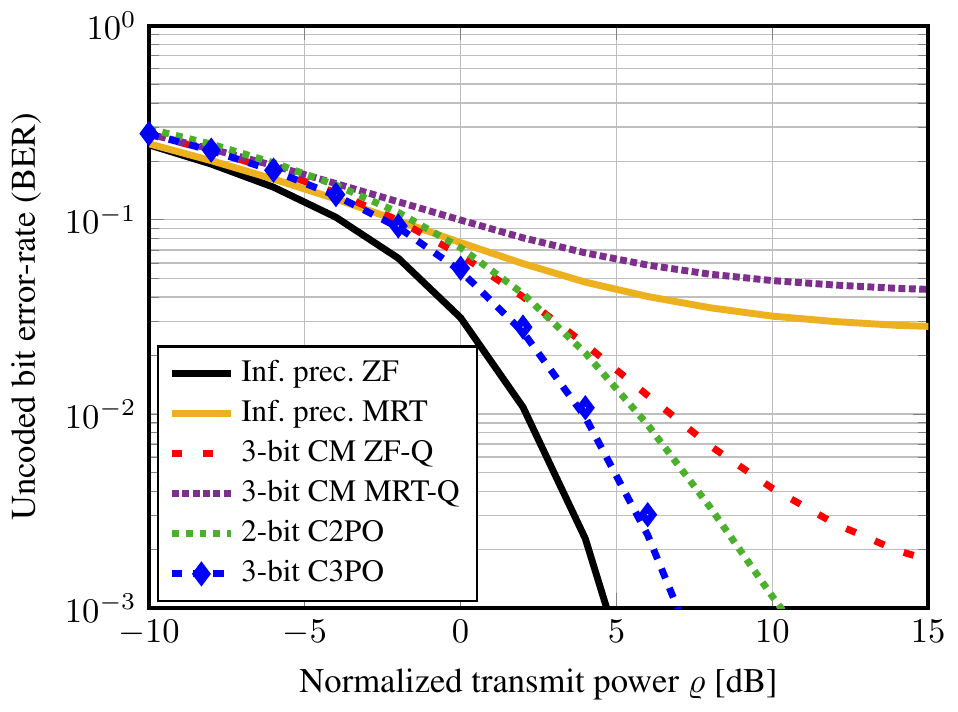}\label{fig:BER_256x16_16QAM}} \quad
\subfigure[Performance/complexity tradeoff.]{\includegraphics[width=0.31\textwidth]{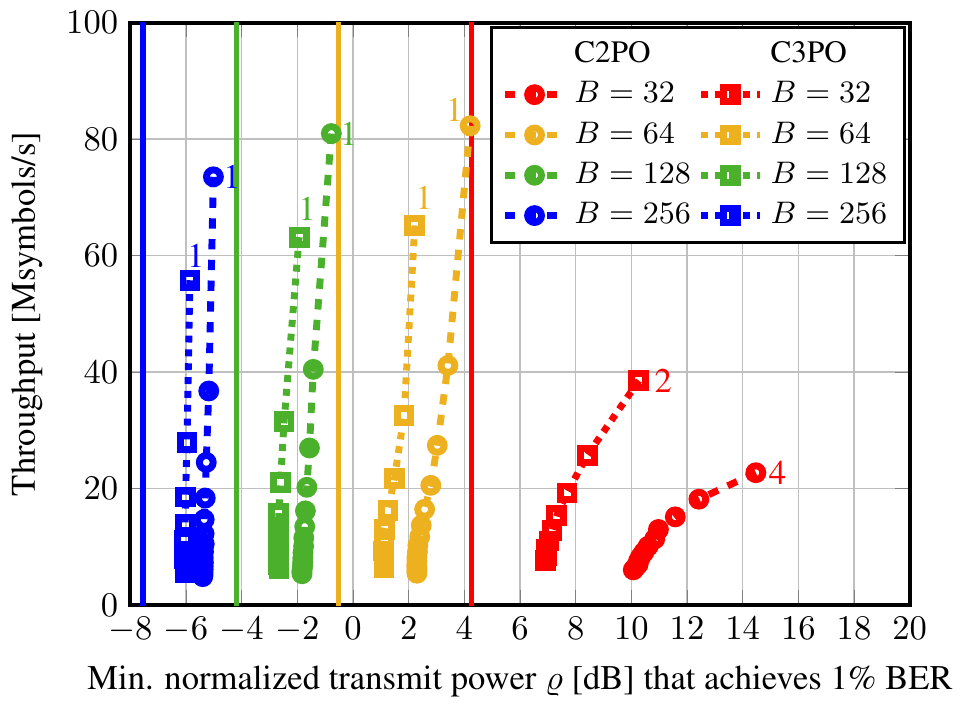}\label{fig:tradeoff}}
\caption{Subfigures (a) and (b): uncoded bit {error-rate} (BER) of various precoders versus normalized transmit power $\snr$.
Markers show fixed-point performance.
Subfigure (c): performance/complexity tradeoffs for C2PO \cite{castaneda2017cxpo} and C3PO; the numbers next to the curves indicate~$t_\text{max}$. The vertical lines show the performance of infinite-precision ZF precoding. C3PO outperforms C2PO in terms of uncoded BER with an increase in implementation complexity.
}
\label{fig:BER_ALL}
\vspace{-3mm}
\end{figure*}

\begin{table*}[tp]
\centering
\renewcommand{\arraystretch}{1.0}
\begin{minipage}[c]{2\columnwidth}
    \begin{center}
    \caption{Xilinx Virtex-7 XC7VX690T~FPGA implementation results for MRT-Q~\cite{castaneda2017cxpo}, C2PO~\cite{castaneda2017cxpo}, and the proposed C3PO for $U=16$ UEs }\vspace{-.2cm}
       \label{tbl:implresultsCxPO}
\scalebox{0.97}{       
  \begin{tabular}{@{}lcccccccccccccc@{}}
  \toprule
  Algorithm & \multicolumn{4}{c}{2-bit CM MRT-Q~\cite{castaneda2017cxpo}}  && \multicolumn{4}{c}{2-bit C2PO~\cite{castaneda2017cxpo}} && \multicolumn{4}{c}{3-bit C3PO (this work) }  \\
  \cmidrule{2-5}  \cmidrule{7-10} \cmidrule{12-15}
  {BS antennas $B$} & $32$ & $64$ & $128$ & $256$  && $32$ & $64$ & $128$ & $256$ && $32$ & $64$ & $128$ & $256$ \\
  \midrule
  {Slices} & 2\,543 & 5\,097 & 9\,444 & 17\,630 && 3\,375 & 6\,519 & 12\,690 & 24\,748 && 8\,765 & 16\,823 & 33\,303 & 65\,451 \\ 
  {LUTs} & 7\,842 & 15\,617 & 32\,476 & 64\,446 && 10\,817 & 21\,920 & 43\,710 & 85\,323 && 29\,034 & 56\,799 & 113\,948 & 224\,420  \\ 
  {Flipflops} & 5\,711 & 11\,419 & 21\,902 & 42\,764 && 5\,677 & 12\,461 & 26\,083 & 53\,409 && 11\,611 & 24\,357 & 49\,893 & 101\,026 \\ 
  {DSP48 units} & 0 & 0 & 0 & 0 && 136 & 272 & 544 & 1\,088 && 136 & 272 & 544 & 1\,088  \\ 
   \midrule   
  Clock frequency [MHz] & 412 & 410 & 388 & 359 && 222 & 206 & 208 & 193 && 202 & 175 & 174 & 157 \\ 
  {Latency$^\textit{a}$ [clock cycles]} & 18 & 18 & 18 & 18 && 39 & 40 & 41 & 42 && 42 & 43 & 44 & 45  \\ 
  {Throughput$^\textit{a}$ [Msymbols/s]} & 366 & 365 & 345 & 319 && 91 & 82 &  81 & 74 && 77 & 65 & 63 & 56  \\ 
  {Power consumption$^\textit{b}$ [W]} & 0.79 & 1.25 & 1.84 & 3.16 && 1.04 & 1.70 & 3.17 & 5.80 && 1.76 & 2.89 & 5.48 & 10.12  \\ 
  \bottomrule
  \end{tabular}
  }
\end{center} 
$^\textit{a}$The {minimum latency and maximum throughput is} measured for one algorithm iteration.\\
\quad $^\textit{b}$Statistical power estimation at maximum clock frequency and 1.0\,V supply voltage.
\end{minipage}
\vspace{-4mm}
\end{table*}

\section{VLSI Architecture and Implementation Results}
\label{sec:architecture}

\subsection{Architecture Overview}

The proposed VLSI architecture is shown in \fref{fig:C3POarch} and builds upon the one of C2PO in \cite{castaneda2017cxpo}, which was designed for 2-bit CM precoding.
As in \cite{castaneda2017cxpo}, we assume that~$B$ is a multiple of $U$, so the architecture consists of $B/U$ linear arrays, each containing $U+1$  processing elements (PEs).
Each linear array operates on a $(U+1)\times U$ sub-matrix of $\overline{\bH}$ and on a $U$-dimensional sub-vector of $\bmx^{(t)}$.
The architecture computes step \fref{eq:faststep1} simplified as in \fref{eq:faststep1-simple} via two separate matrix-vector products using Cannon's algorithm~\cite{CannonThesis}.
We first compute $\bmw = \overline{\bH} (\tau\bmx^{(t)})$ by cyclically exchanging the entries of $\tau\bmx^{(t)}$ between the PEs of the same array.
We then compute $\bmz^{(t+1)} = \bmx^{(t)} - {\tallm}\bmw$ by cyclically exchanging the accumulated results of the PEs within the same array.
Finally, the vector $\bmz^{(t+1)}$ is fed to a projection unit implementing step \fref{eq:faststep2}, thus completing one C3PO iteration.
The proposed architecture requires $2U+\log_2(B/U)+9$ clock cycles for one C3PO iteration.
See \cite{castaneda2017cxpo} for more architecture~details.

Each PE is equipped with (i) an $\tilde{\bmh}_u$ memory storing the 
$u$th row of the corresponding sub-matrix taken from $\overline{\bH}$; (ii) a complex-valued multiply-accumulate (MAC) unit; and (iii) a projection unit.
See \cite{castaneda2017cxpo} for details on (i) and (ii); part (iii), the projection unit, is more complicated than that of C2PO.
Specifically, this unit maps the entries of $\bmz^{(t+1)}$ to the first quadrant of the complex plane and perform comparisons based on the line equations $\ell_1$--$\ell_6$ (see~\fref{sec:proximal}) in order to perform the projection of $\bmz^{(t+1)}$ to~$\setB^B$. 

\subsection{Fixed-Point Parameters}
%
%
The entries of~$\bmx^{(t)}$ use $14$-bit signed values with $8$ fraction bits. The entries of  $\tau \bmx^{(t)}$ use $14$-bit signed values with $13$ fraction bits.
The entries of $\overline{\bH}$ use $11$-bit signed values with 8 fraction bits and are stored in look-up tables (LUTs) used as distributed RAM.
The complex-valued MAC units use $18$-bit signed values with $15$ fraction bits when computing $\bmw$; $11$ fraction bits are used when calculating $\bmz^{(t+1)}$. The adder tree uses $21$ bits with $15$ fraction bits.
The projection unit represents the constants (e.g., $1-\sqrt{2}$ and its reciprocal) using signed values with 4--5 bits, so no multipliers are used in the operations related to lines $\ell_1$--$\ell_6$. A total of $30$ adders and subtractors are used within each projection unit; these components operate signed numbers with $7$ fraction bits; the total bit-width varies between $14$--$15$ bits, depending on the quantity.
%
%

\subsection{Error-Rate Performance}
Fig.~\ref{fig:BER_32x16_BPSK} and Fig.~\ref{fig:BER_256x16_16QAM} show uncoded bit-error rate (BER) as a function of the normalized transmit power $\varrho=B/N_0$ for different precoding algorithms and $U=16$ UEs. \fref{fig:BER_32x16_BPSK} shows the BER for $B=32$ BS antennas and BPSK; \fref{fig:BER_256x16_16QAM},  for $B=256$ BS antennas and 16-QAM.
The simulation results are for $10,000$ Monte-Carlo trials and i.i.d.\ Rayleigh fading channels. Both C2PO and C3PO run with $t_\text{max}=9$.
For reference, we show the BERs with 3-bit CM MRT-quantized (MRT-Q) and ZF-quantized (ZF-Q) precoding, as well as the BERs with MRT (``Inf. prec. MRT'') and ZF precoding (``Inf. prec. ZF'') with infinite-precision DACs.
We see from Fig.~\ref{fig:BER_32x16_BPSK} and Fig.~\ref{fig:BER_256x16_16QAM}  that the nonlinear precoders (C2PO and C3PO) significantly outperform MRT-Q and ZF-Q at high normalized transmit power $\varrho$.
%
%
Furthermore, compared to C2PO, we note that C3PO enables a $3.75$\,dB gain (in terms of $\varrho$) at $1\%$ uncoded BER for $B=32$ and BPSK, and $1.75$\,dB for $B=256$ and 16-QAM.
Finally, we note that the implementation loss of our hardware designs (shown with blue markers) is negligible, i.e., less than~$0.15$\,dB at $1\%$ uncoded BER.

\subsection{FPGA Implementation Results and Comparison} \label{sec:fpgaresults}

Table~\ref{tbl:implresultsCxPO} shows FPGA implementation results for 2-bit CM MRT-Q~\cite{castaneda2017cxpo}, C2PO~\cite{castaneda2017cxpo}, and C3PO.
All designs were developed using Verilog, and implemented using Xilinx Vivado Design Suite for a Xilinx Virtex-7 XC7VX690T FPGA.
The designs support $U = 16$ UEs and were implemented for $B = \{32,64,128,256\}$.
Table~\ref{tbl:implresultsCxPO} reveals that the resources of all designs increase roughly linearly with~$B$.
%
MRT-Q achieves the highest throughput thanks to its simplicity, which comes at the cost of a poor uncoded BER performance.
C2PO uses $\sim\!\!1.4\times$ more LUTs than MRT-Q and requires increased latency and critical path.
Compared to C2PO, C3PO consumes $\sim\!2.6\times$ the number of slices and LUTs, $\sim\!2\times$ the number of flip-flops, and the same number of DSP48s. This difference is caused by the 3-bit CM projection unit, which also increases the latency with its pipeline registers. However, C3PO can significantly outperform C2PO in terms of BER (cf. Fig.~\ref{fig:BER_32x16_BPSK} and Fig.~\ref{fig:BER_256x16_16QAM}).

\subsection{Performance/Complexity Tradeoffs}
Fig.~\ref{fig:tradeoff} shows the performance-complexity tradeoffs of C2PO and C3PO: the complexity is represented by the minimum normalized transmit power $\varrho$ that is required to achieve $1\%$ uncoded BER for BPSK; the performance, by the throughput.
The tradeoffs show systems with BPSK, $U=16$ UEs and $B=\{32,64,128,256\}$ BS antennas. As a reference, the minimum transmit power required for infinite-precision ZF precoding to achieve $1\%$ uncoded BER is shown as a vertical line.
We see from \fref{fig:tradeoff} that, while C2PO is able to achieve higher throughput than C3PO, C3PO requires lower transmit power to achieve $1\%$ uncoded BER.
This difference increases for small array sizes: for a system with $B=32$, 4 iterations of C3PO achieve $1\%$ uncoded BER at $\varrho=8$ dB while C2PO is unable to achieve $1\%$ uncoded BER at such value of~$\varrho$.
\section{Conclusions}
\label{sec:conclusions}

We have proposed a nonlinear precoder for 8-phase (3-bit) CM transmission, C3PO, which builds upon the 4-phase C2PO precoder \cite{castaneda2017cxpo}.
By using a different projection unit and no more than $2.7\times$ higher FPGA resources, C3PO achieves up to $3.75$ dB transmit power reduction, and thus, low uncoded BERs in scenarios for which C2PO exhibits poor error-rate performance.

%
%

%
%
%
\bibliographystyle{IEEEtran}
\bibliography{IEEEabrv,confs-jrnls,publishers,studer,svenbib,tom}


\end{document}